\documentclass[amsmath,amssymb,nofootinbib]{revtex4}
\usepackage{graphicx}
\usepackage{dcolumn}
\usepackage{color}
\usepackage{bm}
\usepackage{amsmath,amssymb,graphicx}
\usepackage{epsfig}
\usepackage{enumerate}
\usepackage{array}
\usepackage{multirow}

\begin{document}
\title
{Quantum dissipation and the virial theorem}
\author{Aritra Ghosh\footnote{ag34@iitbbs.ac.in} and Malay Bandyopadhyay\footnote{malay@iitbbs.ac.in}}
\affiliation{School of Basic Sciences,\\ Indian Institute of Technology Bhubaneswar, Argul, Jatni, Khurda, Odisha 752050, India}
\vskip-2.8cm
\date{\today}
\vskip-0.9cm


\begin{abstract}
In this note, we study the celebrated virial theorem for dissipative systems, both classical and quantum. The classical formulation is discussed and an intriguing effect of the random force (noise) is made explicit in the context of the virial theorem. Subsequently, we derive a generalized virial theorem for a dissipative quantum oscillator, i.e. a quantum oscillator coupled with a quantum heat bath. Such a heat bath is modeled as an infinite collection of independent quantum oscillators with a certain distribution of initial conditions. In this situation, the non-Markovian nature of the quantum noise leads to novel bath-induced terms in the virial theorem. We also consider the case of an electrical circuit with thermal noise and analyze the role of non-Markovian noise in the context of the virial theorem. 
\end{abstract}

\maketitle

\section{Introduction}
In 1870, Rudolf Clausius formulated the virial theorem for a mechanical system \cite{vt1}. For a one-particle system, it says that if the virial \(G = m \mathbf{r}\cdot \mathbf{\dot{r}}\) remains bounded in its time evolution, then 
\begin{equation}\label{1st}
\langle K \rangle_t = -\frac{1}{2} \langle \mathbf{r} \cdot \mathbf{F} \rangle_t,
\end{equation}
where \(K\) is the kinetic energy, \(\mathbf{r}\) is the position vector, and \(\mathbf{F}\) is the external force. Here, angled brackets \(\langle \cdot \rangle_t\) denote `time averaging'. Thus, the virial theorem relates the time-averaged kinetic energy to a suitable time average involving the external force. In particular, if the force is conservative, one has \(\mathbf{F} = -\nabla V\) and thus
\begin{equation}\label{2nd}
\langle K \rangle_t = \frac{1}{2} \langle \mathbf{r} \cdot \nabla V \rangle_t.
\end{equation}
The result can be appropriately generalized to a system of particles. It has found applications in astrophysics, cosmology, molecular physics, quantum mechanics, and statistical mechanics (see for instance \cite{vt2,vt3,vt4,vt5,hv}). \\

The virial theorem can be straightforwardly extended to quantum mechanical systems \cite{vt5,hv} where the Heisenberg equations replace the classical Hamilton's equations of motion. For a quantum mechanical system whose Hamiltonian operator is 
\begin{equation}
H = \frac{p^2}{2m} + V(x),
\end{equation} where \(x\) and \(p\) are the position and momentum operators respectively, i.e. \([x,p] = i \hbar\), the quantum mechanical virial theorem gives 
\begin{equation}\label{qvir}
\bigg\langle \frac{p^2}{m} \bigg\rangle_t =  \langle x V'(x) \rangle_t .
\end{equation} This is in exact analogy with Eq. (\ref{2nd}). Eq. (\ref{qvir}) can be easily generalized to multi-particle systems. In particular, for the harmonic oscillator potential, i.e. for \(V(x) = m \omega_0^2 x^2/2\), one finds that the time-averaged kinetic and potential energies are equal. \\

Since in real life, one rarely has truly isolated systems, one can now ask how Eq. (\ref{qvir}) changes if the quantum system interacts with some environment. As we will discuss in the next section, the corresponding classical result remains unchanged even in the presence of an environment. However, in the quantum mechanical case, we shall find that Eq. (\ref{qvir}) picks up novel environment-induced terms which are non-zero in general. An immediate consequence of this is that the time-averaged kinetic and potential energies of a quantum oscillator are no longer equal. The aim of the present work is to obtain a generalized virial theorem for a dissipative quantum oscillator and to analyze the environment-induced terms in the virial theorem. For that we consider a prototypical system, which consists of a single quantum particle placed in a harmonic potential while it is also interacting with a quantum heat bath. The heat bath is taken to be composed of an infinitely many independent quantum oscillators \cite{FV,CL,lho1}, and the reduced quantum dynamics of the system is described by the quantum Langevin equation \cite{ford1988}. We shall also discuss the result obtained here in the light of the quantum counterpart of energy equipartition theorem studied earlier \cite{jarzy1,jarzy2,jarzy3,jarzy4,jarzy5,jarzy6,kaur,kaur1,kaur2,kaur3}. \\

The paper is organized as follows. Section-(\ref{sec2}) is mostly introductory where we re-derive the classical virial theorem and discuss two simple classical dissipative systems. We emphasize upon the role of environment-induced noise in this context. Then, in section-(\ref{sec3}), we derive a generalized virial theorem for a dissipative quantum oscillator, i.e. for a quantum oscillator interacting with a heat bath. We compute the bath-induced terms in the virial theorem and show that they consistently go to zero as \(\hbar \rightarrow 0\). In section-(\ref{elec}), we consider a linear electrical circuit with thermal noise of quantum origin and analyze the role of non-Markovian noise in the context of the virial theorem. We end with some discussion in section-(\ref{discuss}).

\section{Virial theorem for classical dissipative systems}\label{sec2}
The virial theorem is typically derived in the following way. Let \(H(q^i,p_i)\) describe a Hamiltonian system, with coordinates \(q^i\) and momenta \(p_i\). Then, one defines \(G := q^i p_i\) (repeated indices summed over), which is called the virial. Now, the dynamics of \(G\) is simply 
\begin{equation}
\frac{dG}{dt} = \{G, H\},
\end{equation} where \(\{\cdot,\cdot\}\) is the Poisson bracket. One can then perform time averaging on both sides, i.e. write
\begin{equation}
\lim_{\mathcal{T} \rightarrow \infty}\frac{1}{\mathcal{T}} \int_0^\mathcal{T} \frac{dG}{dt} dt =\lim_{\mathcal{T} \rightarrow \infty}\frac{1}{\mathcal{T}} \int_0^\mathcal{T} \{G, H\} dt.
\end{equation} If \(G\) remains bounded in its time evolution, then one gets the simple result
\begin{equation}\label{HV}
\langle \{G, H\} \rangle_t = 0,
\end{equation} where angled brackets \(\langle \cdot \rangle_t\) denote time averaging. Eq. (\ref{HV}) has been referred to as a hypervirial theorem, or simply a generalized virial theorem \cite{hv} (see also \cite{hvr,virial1,contact}). For a typical conservative mechanical system, one has \(H(q^i,p_i) = K(p_i) + V(q^i)\), where \(K = \sum_i p_i^2/2m_i\) and \(V\) are the kinetic and potential energies, respectively. Thus, Eq. (\ref{HV}) gives
\begin{equation}
\langle \{G, K\} \rangle_t = - \langle \{G, V\} \rangle_t,
\end{equation} which using the fact that \(G = q^i p_i\) and the Leibniz rule for the Poisson bracket leads to
\begin{equation}
\bigg\langle \sum_i \frac{p_i^2}{m_i} \bigg\rangle_t =  \bigg\langle \sum_i q_i \frac{\partial V(q^i)}{\partial q^i} \bigg\rangle_t .
\end{equation} This agrees with Eq. (\ref{1st}) obtained by Clausius, for a conservative system. Although, Eq. (\ref{HV}) only describes conservative systems, it has been extended to the case of contact Hamiltonian dynamics, which describe simple dissipative systems in \cite{contact}. Let us consider two simple dissipative systems below. 

\subsection{Damped oscillator}\label{dampedsec}
The damped oscillator is associated with the following equation of motion:
\begin{equation}\label{dampedeqn1}
m\ddot{x} + \mu \dot{x} + kx = 0,
\end{equation} where symbols have their usual meanings. Thus, the force acting on the particle of mass \(m\) is \(F_x = - \mu \dot{x} - k x\). Since this force is not conservative (for \(\mu \neq 0\)), we cannot apply Eq. (\ref{HV}) to this problem. However, Eq. (\ref{1st}) gives
\begin{equation}\label{dampedlho1}
\langle m \dot{x}^2\rangle_t = \langle k x^2\rangle_t + \mu \langle x \dot{x} \rangle_t,
\end{equation} or equivalently, 
\begin{equation}\label{dampedlho2}
\bigg\langle \frac{ p^2}{m} \bigg\rangle_t = \langle m \omega_0^2 x^2 \rangle_t + \gamma \langle x p \rangle_t,
\end{equation} where \(k = m \omega_0^2\) and \(\mu = m \gamma\). Eq. (\ref{dampedlho2}) is the same as that obtained via a generalized virial theorem in the contact Hamiltonian framework in \cite{contact}. As we shall show in the next subsection, this result [Eq. (\ref{dampedlho1})] differs significantly from that for a Brownian particle where the equation of motion contains a random force (noise) term.

\subsection{Brownian oscillator}\label{cbo}
Consider the case of a Brownian particle suspended in a fluid and moving in a harmonic potential. Then, the equation of motion is the Langevin equation:
\begin{equation}\label{classlang}
m\ddot{x} + \mu \dot{x} + m \omega_0^2 x = F(t),
\end{equation}  where \(F(t)\) is a Gaussian noise, whose power spectrum is white, i.e. 
\begin{equation}
\langle F(t) F(t') \rangle_{\rm th} = \Gamma \delta(t-t'),
\end{equation} for some \(\Gamma > 0\). The Gaussian nature of the noise ensures that all odd moments vanish, i.e. one also has \(\langle F(t) \rangle_{\rm th} =0\), and similarly for other odd moments. It should be noticed that the averaging \(\langle \cdot \rangle_{\rm th}\) which appears above is an averaging over the noise ensemble, i.e. over all possible noise realizations. It will also be referred to as a thermal average, because the noise is associated with thermal properties \cite{Zwanzig,balki}. However, since time averaging of a function which is not periodic involves averaging over a large (infinite) amount of time, in the present case, it turns out that time averages are the same as thermal averages or noise averages. This comes from the necessary assumption of ergodicity or `mixing' which requires that all points in the phase space are explored over an infinitely long time. Henceforth, time averaging over an infinitely long time implies averaging over all possible noise realizations, i.e. over the noise ensemble. In subsequent discussions, we shall be dealing with systems with built-in noise terms, meaning that time averages shall coincide with thermal averages, and vice versa. Thus, for the sake of brevity, we shall drop the subscript from the angled brackets that denote averaging and simply use \(\langle \cdot \rangle\).\\

 Since the net external force on the particle is \(F_x = - \mu \dot{x} - m \omega_0^2 x + F(t)\), from Eq. (\ref{1st}), we get the following result:
\begin{equation}\label{classbrownvir}
\langle m \dot{x}^2\rangle =\langle m \omega_0^2 x^2 \rangle + \mu \langle x \dot{x} \rangle - \langle x F(t) \rangle.
\end{equation}
It is straightforward to show by solving Eq. (\ref{classlang}) that \(\langle x F(t) \rangle = 0\). Now, \( \langle x \dot{x} \rangle\) is the position-velocity correlation function and in the steady state, this too vanishes (see appendix-(\ref{appA})). Therefore, Eq. (\ref{classbrownvir}) simply gives 
\begin{equation}\label{bvir}
\bigg\langle \frac{m \dot{x}^2}{2} \bigg\rangle = \bigg\langle \frac{m \omega_0^2 x^2}{2} \bigg\rangle,
\end{equation} meaning that the averaged kinetic and potential energies are equal. This is in sharp contrast to Eq. (\ref{dampedlho1}), where \(\langle x \dot{x} \rangle_t \neq 0 \) (appendix-(\ref{appA})). Thus, noise terms can essentially control the form of the virial theorem. \\

The presence of a noise term also introduces a thermodynamic interpretation to the averages. For instance, one can solve Eq. (\ref{classlang}) and compute \(\langle x^2 \rangle\) and \(\langle \dot{x}^2 \rangle\) explicitly, by averaging over the noise ensemble. They are found to be equal and read \cite{Zwanzig,balki}
\begin{equation}\label{abcde11}
\bigg\langle \frac{m \dot{x}^2}{2} \bigg\rangle = \bigg\langle \frac{m \omega_0^2 x^2}{2} \bigg\rangle = \frac{\Gamma}{4\mu}.
\end{equation}
Let us now note that the fluid in which the Brownian particle is immersed in has a temperature \(T\) and the particles are distributed according to the Maxwell's speed distribution (we neglect interparticle interactions other than collisions). After repeated collisions, i.e. by exchanging energy with the fluid, the Brownian particle shall also reach a steady state, attaining the same temperature \(T\). Thus, it is described by the canonical distribution:
\begin{equation}
\rho = \frac{e^{-\beta\big(\frac{m\dot{x}^2}{2} + \frac{m \omega_0^2 x^2}{2}\big)}}{Z},
\end{equation} where \(\beta = 1/k_B T\) and \(Z = (2 \pi k_B T/ m \omega_0)\) is the partition function normalizing \(\rho\). The thermally-averaged kinetic energy reads
\begin{equation}
K_{\rm th} =  Z^{-1}\int dx d\dot{x} \bigg(\frac{m \dot{x}^2}{2}\bigg)  e^{-\beta\big(\frac{m\dot{x}^2}{2} + \frac{m \omega_0^2 x^2}{2}\big)} = \frac{k_B T}{2} ,
\end{equation} and similarly \(V_{\rm th} = \frac{k_B T}{2}\). Thus, the thermally-averaged kinetic and potential energies are equal. In the steady state, the time-averaged values agree with the thermally-averaged values. This requires that we pick \(\Gamma = 2 \mu k_B T\) and then everything fits consistently. \\

To reiterate, the presence of a noise term in the equation of motion can dramatically alter the virial theorem. We list the two main features of the noise: 
\begin{enumerate}
\item The presence of noise leads to the averaged kinetic and potential energies to be equal, for the potential \(V(x) = m \omega_0^2 x^2/2\). If the noise was absent, then the time-averaged kinetic and potential energies are not equal [Eq. (\ref{dampedlho1})].

\item The noise associates a natural thermal character to the averages, i.e. the time averages are the same as thermal averages. While Eq. (\ref{dampedlho1}) has no thermal interpretation, Eq. (\ref{bvir}) has one, and the averages are equal to \(\frac{k_B T}{2}\) in accordance with the equipartition theorem.
\end{enumerate}

In the next section, we shall describe a generalized virial theorem for a dissipative quantum oscillator based on a microscopic model approach.  \\

\section{Generalization to quantum dissipation}\label{sec3}
We shall now consider formulating a quantum mechanical virial theorem suited for dissipative systems. A dissipative system is one that interacts with an environment, i.e. the total Hamiltonian is \(H = H_S + H_B + H_{SB}\), where \(H_S\) and \(H_B\) are the Hamiltonians of the system and the bath respectively, while \(H_{SB}\) is that describing the interaction between the system and the bath. However, one is only interested in the dynamics of the system while it is in contact with the heat bath. To obtain that, one conveniently traces over the degrees of freedom of the bath so that one gets a reduced dynamical description for the system and the time evolution is no longer unitary. One way to trace over the bath degrees of freedom is to consider the path integral formulation of quantum mechanics and subsequently compute the influence functional which takes care of the environmental effects \cite{FV}. A distinct approach is to consider the Heisenberg equations for the system and bath variables, wherein one solves those for the bath variables and substitutes them into the equations of motion describing the system degrees of freedom. This naturally leads to a quantum Langevin equation \cite{ford1988}, with a built-in fluctuation-dissipation theorem. \\

In this paper, we consider the latter approach to dissipative quantum systems. Our system is a particle of mass \(m\), moving in one dimension in the presence of a harmonic potential \(V(x) = m \omega_0^2 x^2/2\). The bath is modeled as a collection of infinitely many quantum oscillators with a distribution of initial conditions, while the system-bath coupling is taken to be bilinear. Thus, the total Hamiltonian reads 
\begin{equation}\label{H}
H = \frac{p^2}{2m} + V(x) + \sum_{j = 1}^N \Bigg[ \frac{p_j^2}{2m_j} + \frac{m_j \omega_j^2}{2} \big(q_j - x \big)^2 \Bigg],
\end{equation} where \(q_j\) and \(p_j\) are the coordinates and momenta of the bath degrees of freedom, i.e. \([q_j,p_j] = i \hbar \delta_{ij}\). It should be remarked that we have considered an additional potential term \(V_r(x) = x^2 \sum_{j = 1}^N m_j \omega_j^2\), in order to ensure that the system-bath coupling is homogeneous in space. Addition of \(V_r(x)\) is often called `potential renormalization'. \\

Now, one can use the usual commutation relations between the variables appearing in Eq. (\ref{H}) to derive the Heisenberg equations for the system and bath variables. Subsequently, one can solve the equations of motion for the bath variables and substitute them into the equation of motion describing the system. The resulting reduced equation of motion describing the `open' quantum system is
\begin{equation}\label{qle}
m \ddot{x} + \int_{-\infty}^t \mu(t - t') \dot{x}(t') dt' + V'(x) = F(t),
\end{equation} where the function \(\mu(t)\) is the dissipation kernel whose expression is 
\begin{equation}\label{1}
\mu(t) = \Theta(t) \sum_{j = 1}^N m_j \omega_j^2 \cos (\omega_j t) .
\end{equation} The presence of the step function \(\Theta(t)\) ensures that \(\mu(t)\) vanishes for negative arguments, consistent with the principle of causality. Now, assuming that the system and the bath were at equilibrium at the initial instant, \(F(t)\) is a Gaussian noise whose explicit expression is
 \begin{equation}\label{noise}
 F(t)=\sum_{j=1}^{N} m_j \omega_j^2 \Bigg[\Big(q_j(0)-x(0)\Big)\cos(\omega_jt)+\frac{p_j(0)}{m_j\omega_j}\sin(\omega_jt)\Bigg].
 \end{equation}
Notice that the noise depends upon the initial conditions of the bath oscillators, as well as that of the system. The initial conditions can be taken to be distributed according to a canonical distribution, and this implements the random nature of \(F(t)\), lending a thermal character to the averages performed over the noise ensemble. This is not a quantum mechanical feature and the same arguments go through for the classical case. Furthermore, it should also be noted that the thermodynamic limit corresponds to \(N \rightarrow \infty\), which shall be assumed to be the case even if it is not explicitly mentioned. The spectral properties of the noise are given by the following correlation function and commutator (see for instance \cite{purisdgbook}, chapter 9):
\begin{eqnarray}
\langle \lbrace F(t), F(t') \rbrace \rangle &=& \frac{2}{\pi}\int_{0}^{\infty}d\omega \hbar \omega {\rm Re} [ \tilde{\mu} (\omega)] \coth\Big(\frac{\hbar\omega}{2k_BT}\Big) \cos \lbrack \omega(t-t')\rbrack,  \label{symmetricnoisecorrelation1} \\
\langle \lbrack F(t), F(t') \rbrack \rangle &=& \frac{2}{i\pi}\int_{0}^{\infty}d\omega \hbar \omega   {\rm Re}[ \tilde{\mu} (\omega)] \sin\lbrack \omega(t-t')\rbrack, \label{noisecommutator1}
\end{eqnarray} where \(\tilde{\mu}(\omega)\) is the Fourier transform of \(\mu(t)\) [Eq. (\ref{tildemu})]. \\

 It is often convenient to define a bath spectral function \(J(\omega)\) describing the spectral distribution of the bath degrees of freedom as 
\begin{equation}\label{2}
J(\omega) = \frac{\pi}{2} \sum_{j=1}^N m_j \omega_j^3 \delta(\omega - \omega_j).
\end{equation} Combining Eqs. (\ref{1}) and (\ref{2}), one gets 
\begin{equation}\label{mu}
\mu(t) = \frac{2}{\pi} \int_0^\infty \frac{J(\omega)}{\omega} \cos (\omega t) d\omega.
\end{equation}
We have now set up our notation and have introduced all the preliminaries needed to derive a generalized virial theorem for the dissipative quantum oscillator. \\

Now let us derive a generalized virial theorem for the dissipative quantum system described above. Since we are interesting in the system (\(S\)), we define a symmetrized virial operator for the system as \(G = \frac{xp + px}{2} \), which means  
\begin{equation}\label{dGdt}
\frac{d\langle G \rangle}{dt} = \frac{\langle [G,H] \rangle}{i\hbar} =  \frac{\langle [xp,H] + [px,H] \rangle}{2i\hbar}.
\end{equation}
Note that the Hamiltonian \(H\) appearing in the equation above is the total Hamiltonian [Eq. (\ref{H})] which also contains \(H_B\) and \(H_{SB}\). Let us analyze the quantity \( \langle [xp,H] + [px,H] \rangle \). Some straightforward manipulations using Eq. (\ref{H}) give
\begin{widetext}
\begin{eqnarray}
 \langle [xp,H] + [px,H] \rangle & =&\bigg\langle \frac{p^2}{m} \bigg\rangle -\big\langle x V'(x) \big\rangle + \sum_{j=1}^N m_j \omega_j^2 \langle q_j(t) x(t) \rangle - \frac{1}{2} \sum_{j=1}^N m_j \omega_j^2 \int_{-\infty}^t dt' \langle x(t) \dot{x}(t') + \dot{x}(t') x(t) \rangle \cos [\omega_j (t - t')]. \nonumber \\
\end{eqnarray}
Let us define 
\begin{eqnarray}
I_1 &=& \frac{1}{2}  \sum_{j=1}^N m_j \omega_j^2 \langle q_j(t) x(t) + x(t) q_j(t)  \rangle = \frac{1}{2} \langle x(t) F(t) + F(t) x(t) \rangle, \label{i1} \\
I_2 &=& \frac{1}{2} \sum_{j=1}^N m_j \omega_j^2 \int_{-\infty}^t dt' \langle x(t) v(t') + v(t') x(t) \rangle \cos [\omega_j (t - t')] =  \int_{-\infty}^t dt' \mu(t-t') C_{xv} (t-t'), 
\end{eqnarray} 
\end{widetext} where we have used Eqs. (\ref{noise}) and (\ref{mu}), and have defined the position-velocity correlation function \(C_{x v} (t-t') = \frac{1}{2}  \langle x(t) v(t') + v(t') x(t) \rangle\), for the velocity operator \(v(t) = \dot{x}(t)\).\\

If one now takes the steady-state limit, one would have \(d\langle G \rangle/dt = 0\), meaning that the left-hand side of the equation above vanishes. This will give Eq. (\ref{qvir}) together with novel bath-induced terms which can be determined by evaluating the steady-state expressions for \(I_1\) and \(I_2\). One may interpret here, \(c_j = m_j \omega_j^2\) as coupling constants which couple the system and the environment. Setting \(c_j = 0\), one straightforwardly recovers Eq. (\ref{qvir}), as anticipated. In the steady state, for \(V(x) = m \omega_0^2 x^2/2\), one has \(\langle K \rangle - \langle V \rangle = \frac{I_2 - I_1}{2}\), and therefore the averaged kinetic and potential energies are unequal in general \cite{jarzy3,jarzy4,kaur}. We now separately analyze \(I_1\) and \(I_2\). 

\subsection{Analysis of \(I_1\)}
Let us begin by evaluating \(I_1\) defined in Eq. (\ref{i1}). Physically, this term corresponds to the average work done on the dissipative oscillator by the quantum noise. Remarkably, it vanishes for the classical case \cite{classnoise} but as we will show, it is non-trivial for the dissipative quantum oscillator. Now, \(I_1\) is the symmetrized position-noise correlation function, i.e. 
\begin{equation}\label{I11}
I_1 = C_{xF} := \frac{\langle x(t) F(t) +   F(t) x(t) \rangle}{2},
\end{equation}
and we are required to compute the above correlation function for equal times, i.e. the position and noise operators are defined at the same time instant \(t\). For the moment however, let us consider a more general unequal time correlation function, where \(\tau = t -t'\) is the time separation, i.e. we consider \(C_{xF}(t-t') := \frac{\langle x(t) F(t') +   F(t') x(t) \rangle}{2}\), and later we will set \(t =t'\) or \(\tau = 0\). Now, we may write
\begin{equation}\label{ggg}
C_{xF}(\tau) = \frac{1}{2 \pi} \int_{-\infty}^\infty d\omega \tilde{C}_{xF} (\omega) e^{- i \omega \tau},
\end{equation} which means if we can find \( \tilde{C}_{xF} (\omega)\), we can find \(C_{xF}(\tau)\) at \(\tau = 0\) easily. Let us begin by considering the quantum Langevin equation [Eq. (\ref{qle})] with \(V(x) = m \omega_0^2 x^2/2\). A Fourier transform allows us to solve it and we may write \(\tilde{x} (\omega) = \alpha (\omega) \tilde{F} (\omega)\), where
\begin{equation}
\tilde{x}(\omega) = \int_{-\infty}^\infty x(t) e^{i \omega t} dt, \hspace{4mm} \tilde{F}(\omega) = \int_{-\infty}^\infty F(t) e^{i \omega t} dt,
\end{equation} and
\begin{equation}\label{alphadef}
\alpha(\omega) = \frac{1}{m (\omega_0^2 - \omega^2) - i \omega \tilde{\mu}(\omega)}
\end{equation} is the susceptibility. In the preceding expression, 
\begin{equation}\label{tildemu}
\tilde{\mu}(\omega) = \int_0^\infty \mu(t) e^{i \omega t}
\end{equation}\\ is the Fourier-transformed dissipation kernel. Since, \(\tilde{x} (\omega) = \alpha (\omega) \tilde{F} (\omega)\), one can write 
\begin{equation}
\tilde{C}_{xF} (\omega) = \alpha (\omega) \tilde{C}_{FF} (\omega),
\end{equation} where \(\tilde{C}_{FF} (\omega) = \int_{-\infty}^\infty C_{FF}(\tau) e^{ i \omega \tau}\) is the Fourier-transformed noise autocorrelation function, which reads
\begin{equation}
\tilde{C}_{FF} (\omega) = {\rm Re} [\tilde{\mu}(\omega)] \hbar \omega \coth \bigg(\frac{\hbar \omega}{2 k_B T}\bigg).
\end{equation} This means from Eqs. (\ref{I11}) and (\ref{ggg}), we have
\begin{equation}
I_1 = \frac{\hbar}{2 \pi} \int_{-\infty}^\infty d\omega \omega  {\rm Re} [\tilde{\mu}(\omega)] \alpha(\omega) \coth \bigg(\frac{\hbar \omega}{2 k_B T}\bigg),
\end{equation} where we have set \(\tau = 0\) in Eq. (\ref{ggg}). The reader should note that \(I_1\) vanishes in the weak-coupling limit, for which \(\tilde{\mu}(\omega) \rightarrow 0\). Although so far we have kept our analysis general, for ease of computing the exact expression of \(I_1\), we now resort to Ohmic dissipation for which \(\tilde{\mu}(\omega) = m \gamma\), where \(\gamma > 0\) is the system-bath coupling strength. It should be remarked that although we use this specific dissipation mechanism owing to its simplicity, the Ohmic bath is often considered controversial \cite{jarzy6}, primarily because it leads to divergences of certain quantities, such as the mean kinetic energy, or the potential \(V_r(x)\) introduced earlier for potential renormalization (see also \cite{ohmicdiverge}). For Ohmic dissipation, one simply has 
\begin{eqnarray}
I_1 &=& \frac{\hbar \gamma}{2 \pi} \int_{-\infty}^\infty  \frac{\omega d\omega}{(\omega_0^2 - \omega^2) - i \omega \gamma} \coth \bigg(\frac{\hbar \omega}{2 k_B T}\bigg) \nonumber \\
&=& \frac{\gamma}{ \beta \pi} \int_{-\infty}^\infty  \frac{ d\omega}{(\omega_0^2 - \omega^2) - i \omega \gamma} \bigg(\frac{\hbar \omega}{2 k_B T}\bigg) \coth \bigg(\frac{\hbar \omega}{2 k_B T}\bigg) \nonumber \\
&=&  \frac{\gamma}{ \beta \pi} \int_{-\infty}^\infty  \frac{\omega d\omega}{(\omega_0^2 - \omega^2) - i \omega \gamma}  \Bigg[1 + 2 \sum_{n=1}^\infty \frac{\omega^2}{\omega^2 + \nu_n^2} \Bigg], \label{I1integral1}
\end{eqnarray} where \(\nu_n = \frac{2 \pi n }{\hbar \beta}\) are the bosonic Matsubara frequencies, and we have used the formula \(z \coth z = 1 + 2 \sum_{n=1}^\infty \frac{z^2}{z^2 + (n \pi)^2}\), where \(z\) is a complex argument. While performing the integral, one has to choose among \(\gamma/2 < \omega_0\) (under-damped regime), \(\gamma/2 > \omega_0\) (over-damped regime), or \(\gamma/2 = \omega_0\) (critically-damped case). Let us focus on the over-damped regime, for which, upon evaluating the integral by closing the contour on the lower half-plane, we find
\begin{widetext}
\begin{equation}\label{I1final}
I_1 = \frac{2 \gamma}{\beta} \sum_{n=1}^\infty \frac{\nu_n}{(\nu_n - \omega_+)(\nu_n - \omega_-)} + \frac{4 \gamma}{\beta} \sum_{n=1}^\infty \frac{\omega_+^2}{(\nu_n^2 - \omega_+^2)(\omega_+ - \omega_-)} + \frac{4 \gamma}{\beta} \sum_{n=1}^\infty \frac{\omega_-^2}{(\nu_n^2 - \omega_-^2)(\omega_+ - \omega_-)},
\end{equation}
\end{widetext} where
\begin{equation}
\omega_{\pm} = \frac{\gamma}{2} \pm \sqrt{\frac{\gamma^2}{4} - \omega_0^2}
\end{equation} are real numbers. One can immediately see that \(\hbar \rightarrow 0\) gives \(I_1 \rightarrow 0\). Thus, for the classical Brownian oscillator, the average work done by the random force is exactly zero, whereas, it is non-zero for the quantum case. This seems to originate from the non-Markovian nature of the quantum noise, which becomes Markovian for \(\hbar \rightarrow 0\). 

\subsection{Analysis of \(I_2\)}
Consider now
\begin{equation}\label{I2def11}
I_2 = \int_{-\infty}^t dt' \mu(t-t') C_{xv} (t-t').
\end{equation} Since \(\mu(t)\) signifies the friction memory, it must go to zero for \(t < 0\). Thus, we can replace the upper limit in the integral above with \(+\infty\), which gives 
\begin{eqnarray}
I_2 &=& \int_{-\infty}^\infty dt' \mu(t-t') C_{xv} (t-t')  \nonumber \\
 &=&  \int_{-\infty}^\infty dt' \mu(t-t') \frac{1}{2\pi} \int_{-\infty}^\infty d\omega \tilde{C}_{xv} (\omega) e^{-i\omega(t-t')}. 
 \label{abc}
\end{eqnarray} Let us note that \(\tilde{C}_{xv}(\omega) = i \omega \tilde{C}_{xx}(\omega)\), where
\begin{equation}\label{CW}
\tilde{C}_{xx} (\omega) = \hbar  {\rm Im} [\alpha(\omega)] \coth \bigg(\frac{\hbar \omega}{2 k_B T}\bigg),
\end{equation} by the fluctuation-dissipation theorem \cite{callenwelton,landau}. Then, Eq. (\ref{abc}) can be re-written as (see also \cite{EB})
\begin{eqnarray}
I_2 &=& \frac{i \hbar}{2\pi}  \int_{-\infty}^\infty \omega  {\rm Im} [\alpha(\omega)] \coth \bigg(\frac{\hbar \omega}{2 k_B T}\bigg) \tilde{\mu} (-\omega) d\omega \nonumber \\
&=& \frac{i}{\beta \pi}  \int_{-\infty}^\infty  {\rm Im} [\alpha(\omega)] \bigg(\frac{ \hbar \omega}{2 k_B T}\bigg) \coth \bigg(\frac{\hbar \omega}{2 k_B T}\bigg) \tilde{\mu} (-\omega) d\omega \nonumber \\ 
&=& \frac{i}{\beta \pi}  \int_{-\infty}^\infty  {\rm Im} [\alpha(\omega)] \Bigg[1 + 2 \sum_{n=1}^\infty \frac{\omega^2}{\omega^2 + \nu_n^2} \Bigg] \tilde{\mu} (-\omega) d\omega,  
\label{I2ex111}
\end{eqnarray} where \(\nu_n = \frac{2 \pi n }{\hbar \beta}\) are the bosonic Matsubara frequencies, and we have used the summation formula for \(z \coth z\) for complex \(z\). Notice that Eqs. (\ref{I2def11})-(\ref{I2ex111}) hold for a general \(\tilde{\mu}(\omega)\) describing the heat bath. However, in order to explicitly compute the expression for \(I_2\), we now resort to Ohmic dissipation, i.e. \(\tilde{\mu}(\omega) = m \gamma\), which means \(\alpha (\omega) = [ m (\omega_0^2 - \omega^2) - i m \gamma \omega]^{-1} \). Therefore, we have
\begin{equation}
{\rm Im} [\alpha(\omega)] = \frac{\gamma \omega}{m[ (\omega_0^2 - \omega^2)^2 + (\gamma \omega)^2]}.
\end{equation}
This gives
\begin{equation}
I_2 =  \frac{i \gamma^2}{\beta \pi}  \int_{-\infty}^\infty   \frac{ \omega d\omega}{[ (\omega_0^2 - \omega^2)^2 + (\gamma \omega)^2]} \Bigg[1 + 2 \sum_{n=1}^\infty \frac{\omega^2}{\omega^2 + \nu_n^2} \Bigg]. \end{equation}
The integral can be evaluated on the lower half-plane with some straightforward steps, giving
\begin{eqnarray}
I_2 &=& \frac{\gamma^2}{\beta} \sum_{n=1}^\infty \frac{\nu_n^2}{\big(\omega_0^2 + \nu_n^2  - \frac{\gamma^2}{2} \big)^2 + (\Omega \gamma)^2 } - \frac{\gamma^2}{\beta} \sum_{n=1}^\infty \frac{\nu_n^2}{(\omega_0^2 + \nu_n^2)^2 - (\gamma \nu_n)^2}, \label{I2final}
\end{eqnarray} where
\begin{equation}\label{Omega}
\Omega = \sqrt{\omega_0^2 - \frac{ \gamma^2}{4}}. 
\end{equation}
Let us look at two important limiting cases. First, as one puts \(\hbar \rightarrow 0\), one has \(\nu_n \rightarrow \infty\), for all values of \(n\). Thus, the series converges to \(I_2 = 0\) which is the correct classical limit. Next, we take the limit \(\gamma \rightarrow 0\), which is the weak-coupling regime. In this case, it is easy to verify that \((I_2)_{\gamma \rightarrow 0} = 0\). This is once again consistent with Eq. (\ref{qvir}), because \(I_2\) is an environment-induced term in the virial theorem and goes away if environmental effects are weak. Below, we study the generalized virial theorem for an electrical circuit with non-Markovian thermal noise.  

\section{Electrical circuit with Johnson-Nyquist noise}\label{elec}
In this section, we will consider an \(LCR-\)circuit in the presence of thermal voltage (noise) originating from the thermal motion of electrons. Such a thermal noise is called Johnson-Nyquist noise \cite{johnson1,johnson2,nyquist}, and is present at all non-zero temperatures. For a circuit with no active or non-linear elements, where the \(L\), \(C\), and \(R\) elements are placed in series, Kirchhoff's voltage rule gives 
\begin{equation}\label{KVL}
L \frac{d^2 Q(t)}{dt^2} + R \frac{dQ(t)}{dt} + \frac{Q(t)}{C} = \mathcal{V}(t),
\end{equation} where \(Q (t)\) is the time-varying electric charge and \(\mathcal{V}(t)\) is the time-varying voltage. The time-varying current in the circuit is \(I(t) = \frac{dQ(t)}{dt}\).  Typically, \(\mathcal{V}(t) = V(t) + F(t)\), where \(V(t)\) is the applied voltage (from the source), while \(F(t)\) is the Johnson-Nyquist (thermal) noise arising due to the random motion of electrons at finite temperature. Thus, Eq. (\ref{KVL}) takes the form of a Langevin equation. For simplicity, we take a situation where \(V(t) = 0\), and therefore any small time-varying current in the circuit is entirely due to the thermal voltage. Taking into account the quantum mechanical nature of the thermal modes, the noise is not white but is associated with the power spectrum \cite{johnson1,johnson2,nyquist} (see also \cite{thermo,ghosh}):
\begin{equation}\label{noisepowerspectrum}
S_F (\omega) d \omega = \frac{(2/\pi)R(\omega)\hbar \omega d\omega}{e^{\hbar \omega/k_B T} - 1} ,
\end{equation} where \(R(\omega)\) is the transfer resistance, which is often independent of frequency, i.e. we will put \(R(\omega) = R\). Thus, the thermal modes are distributed according to the black-body distribution. Since the power spectrum explicitly depends upon \(\omega\), the noise is non-Markovian. \\

Let us note that Eq. (\ref{KVL}) is not a quantum Langevin equation in the usual sense [Eq. (\ref{qle})], because \(Q(t)\) is not an operator, and there is no microscopic Hamiltonian such as Eq. (\ref{H}) from which it has been derived. On the other hand, the problem is intrinsically quantum mechanical due to the noise not being white, but rather depending upon the frequency. Therefore, this problem deserves separate attention in the context of the virial theorem. It is easy to check that as \(\hbar \rightarrow 0\), the power spectrum becomes a constant, independent of the frequency, i.e. \(S_F (\omega)\big|_{\hbar \rightarrow 0} = 2 R k_B T/\pi \). \\

In \cite{ghosh}, the mean energies in the capacitor \(C\) and the inductor \(L\) were computed, and it was demonstrated that they are consistent with the quantum counterpart of energy equipartition theorem \cite{jarzy1,jarzy2,jarzy3,jarzy4,jarzy5,jarzy6,kaur,kaur1,kaur2,kaur3}. We refer the reader to \cite{ghosh} for the detailed calculation of the mean energies using spectral analysis, and quote the final expressions below:
\begin{equation}\label{EL}
\langle E_L \rangle = \int_{-\infty}^\infty  \frac{(R \omega^2/\pi L)}{ (\omega^2 - \omega_0^2)^2 + \omega^2 (R/L)^2} \epsilon(\omega, T) d\omega,
\end{equation}
\begin{equation}\label{EC}
\langle E_C \rangle = \int_{-\infty}^\infty  \frac{(R \omega_0^2/\pi L)}{ (\omega^2 - \omega_0^2)^2 + \omega^2 (R/L)^2}  \epsilon(\omega, T) d\omega ,
\end{equation}
where \(\omega_0 = 1/\sqrt{LC}\) and for convenience, we will include the zero-point energy contribution in \(S_F(\omega)d\omega\), which implies that
\begin{equation}
\epsilon(\omega,T) = \frac{\hbar \omega}{4} \coth \bigg(\frac{\hbar \omega}{2 k_B T}\bigg). 
\end{equation} 
Clearly, the mean energies \( \langle E_L \rangle \) and \( \langle E_C \rangle \) of the inductive and capacitive elements respectively, are not equal, because the integrands in Eqs. (\ref{EL}) and (\ref{EC}) are different. However, as \(\hbar \rightarrow 0\), it is easy to verify that \(\langle E_L \rangle = \langle E_C \rangle = \frac{k_B T}{2}\), which is the genuine classical limit for which the noise is white, i.e. Markovian. The fact that in general \(\langle E_L \rangle \neq \langle E_C \rangle\) is the electrical analogue of the statement that the mean kinetic and potential energies of the dissipative oscillator are unequal, which we studied in the previous section. Since at present, there is no microscopic Hamiltonian for the electrical problem from where we can compute a virial theorem, let us analyze the difference:
\begin{widetext}
\begin{eqnarray}
\langle E_L \rangle - \langle E_C \rangle &=& \frac{R}{\pi L} \int_{-\infty}^\infty  \frac{(\omega^2 - \omega_0^2)}{ (\omega^2 - \omega_0^2)^2 + \omega^2 (R/L)^2} \epsilon(\omega, T) d\omega \nonumber \\
&=& \frac{ \gamma}{2 \pi \beta } \int_{-\infty}^\infty  \frac{ \big(\omega^2 - \omega_0^2 \big) d\omega}{ (\omega_0^2 - \omega^2)^2 + (\gamma \omega)^2}   \Bigg[1 + 2 \sum_{n=1}^\infty \frac{\omega^2}{\omega^2 + \nu_n^2} \Bigg], \label{ELECdiff1}
\end{eqnarray} 
where we have put \(\gamma = R/L\) and have used the summation formula for the coth function. This difference would be analogous to the combined contributions of \(I_1\) and \(I_2\) computed for the quantum Brownian oscillator in the previous section. In Eq. (\ref{ELECdiff1}), the first term inside the parenthesis (\(n=0\) contribution) integrates to zero. The remaining terms give non-trivial contributions which are computed to be\footnote{\textbf{Added note:} It should be noted that the series appearing in Eq. (\ref{elecser}) diverges because some terms scale as \(\sim 1/n\), for large \(n\). We should remark that although Eq. (\ref{elecser}) has been obtained by integrating over the upper half-plane, a similar computation on the lower half-plane also leads to a \(1/n\) divergence.}
\begin{eqnarray}
\langle E_L \rangle - \langle E_C \rangle = \frac{\gamma}{\beta} \sum_{n=1}^\infty \frac{\nu_n (\nu_n^2 + \omega_0^2)}{(\nu_n^2 + \omega_0^2)^2 - (\gamma \nu_n)^2} - \frac{2\gamma^2}{\beta}  \sum_{n=1}^\infty \frac{\nu_n^2}{\big(\omega_0^2 + \nu_n^2 - \frac{\gamma^2}{2}\big)^2 + (\gamma \Omega)^2}, \label{elecser}
\end{eqnarray} where \(\Omega\) is defined in Eq. (\ref{Omega}). \\
\end{widetext}
In the classical limit, i.e. for \(\nu_n \rightarrow \infty\) for all \(n\), all the terms in the series appearing in Eq. (\ref{elecser}) become zero, meaning that the mean energies are equal, i.e. \(\langle E_L \rangle = \langle E_C \rangle = \frac{k_B T}{2}\). The same conclusions are obtained in the high-temperature limit, for which \(\beta \rightarrow 0\). It is noteworthy that we had considered the Johnson-Nyquist noise to be a non-Markovian quantum noise with power spectrum given in Eq. (\ref{noisepowerspectrum}). If instead, one considered a Markovian noise, such as that which arises from Eq. (\ref{noisepowerspectrum}) as \(\hbar \rightarrow 0\), one would obtain \(\langle E_L \rangle = \langle E_C \rangle\). Thus, we have demonstrated the validity of the generalized virial theorem in the case of a linear circuit with thermal noise.

\section{Discussion}\label{discuss}
In this paper, we have discussed the virial theorem for both classical and quantum dissipative oscillators. We demonstrated the crucial role played by noise in this context. The presence of noise lends a thermal character to the averages, i.e.~the time averages coincide with thermal averages. Further, the Markovian nature of the classical noise ensures that the averaged kinetic and potential energies of an oscillator are equal. On the other hand, for a quantum system, the noise is non-Markovian and thus, apart from lending a thermal nature to the averages, it leads to novel dissipation-induced terms in the virial theorem. For the
dissipative quantum oscillator, the generalized virial theorem acquires two additional
contributions due to the heat bath. The first one, which is denoted by
$I_1$, is the average work done on the system by the non-Markovian
quantum noise. The other contribution to the generalized virial theorem is
denoted by $I_2$, whose final expression in the form of an infinite
series has been given in Eq. (\ref{I2final}). Following the analysis presented
in~\cite{EB}, we believe that it is related to the average rate of power
transfer due to dissipation. The kinetic and potential energies are no longer
equal, except in the classical or (quantum) weak-coupling limit, for which the
bath-induced corrections vanish. \\

\begin{figure}
\begin{center}
\includegraphics[scale=0.55]{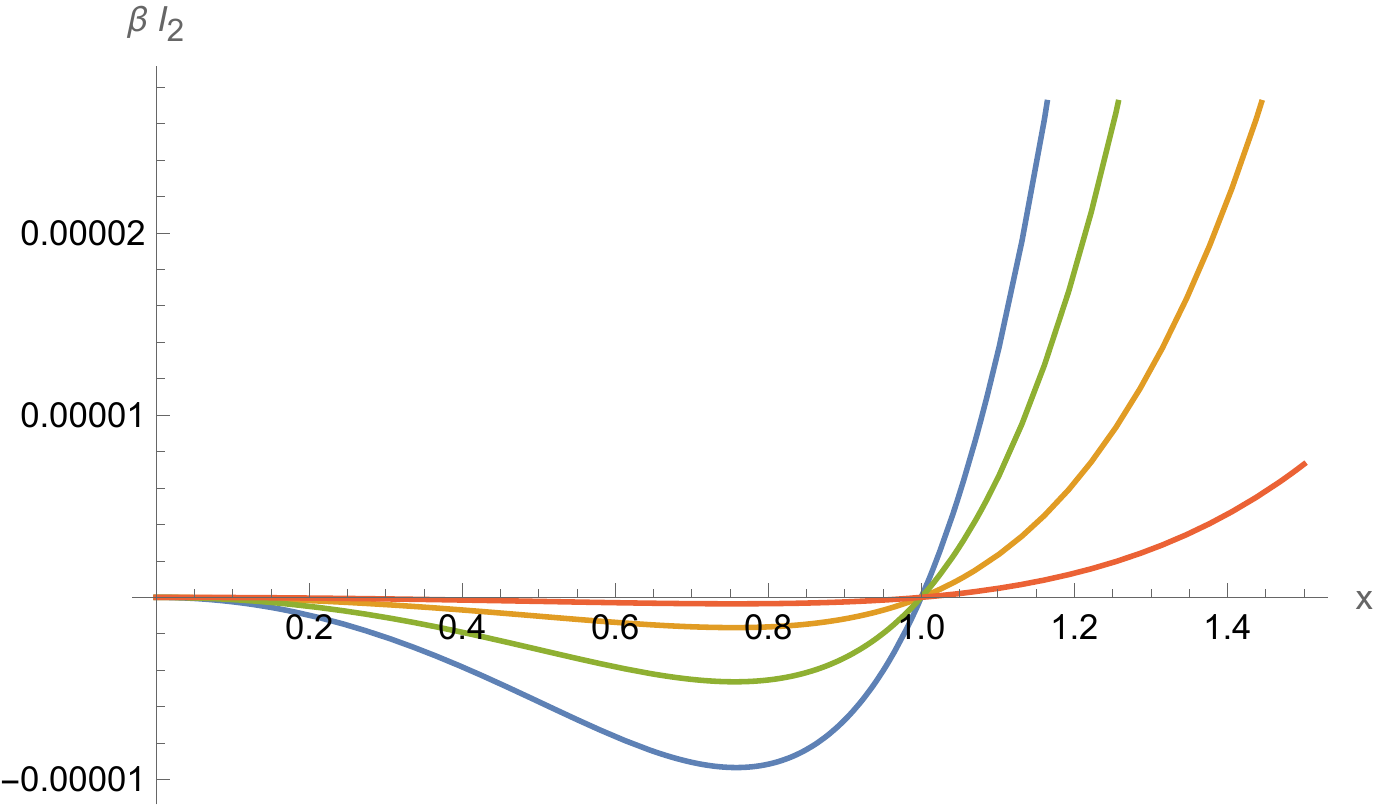}
\caption{Plot of \(\beta I_2\) as a function of \(x = \beta \hbar \omega_0\), for \(\rho = \gamma \omega_0^{-1}\) = \(0.50\) (red), \(0.75\) (yellow), \(1.00\) (green), and \(1.25\) (blue). We have summed over the first 500 terms of the series appearing in Eq. (\ref{I2final1}).}
\label{fig111}
\end{center}
\end{figure}

\begin{figure}
\begin{center}
\includegraphics[scale=0.55]{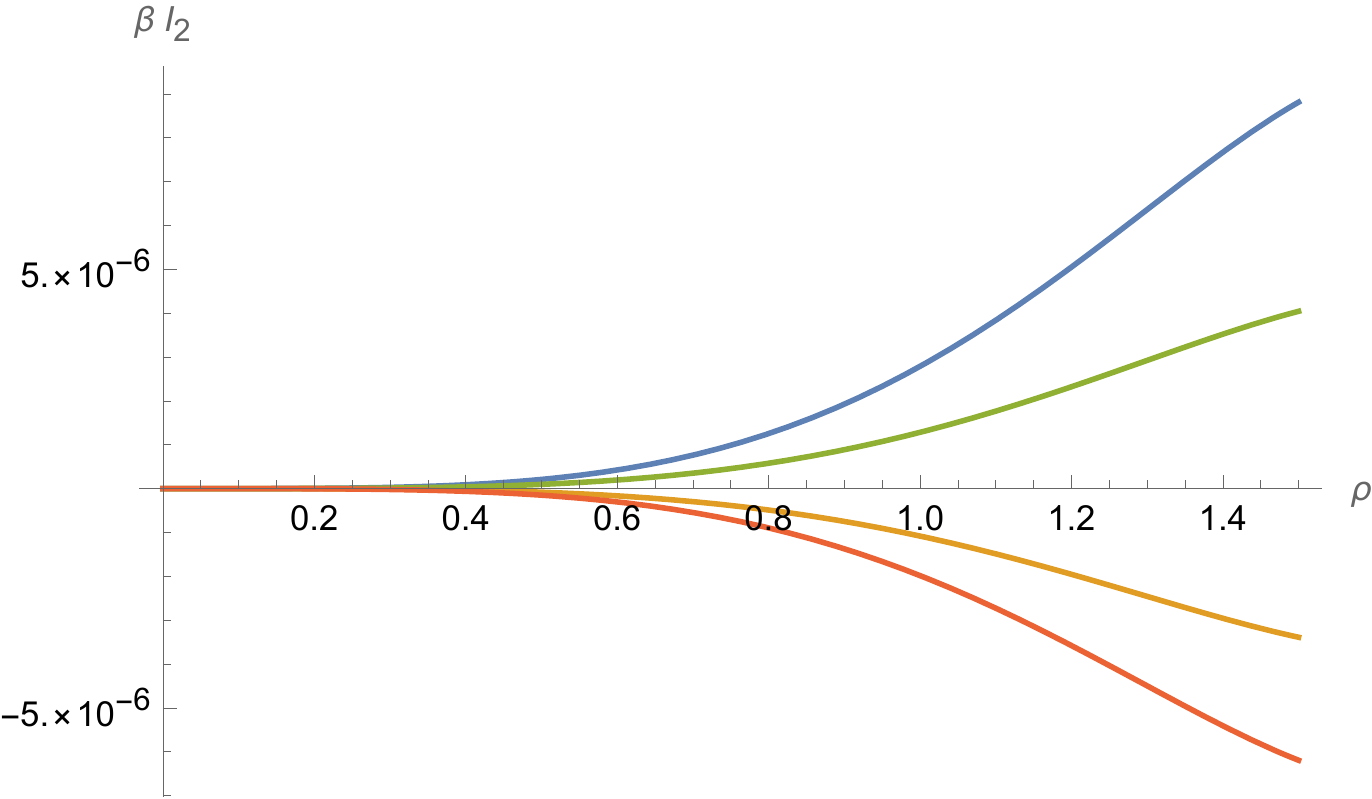}
\caption{Plot of \(\beta I_2\) as a function of \(\rho = \gamma \omega_0^{-1}\), for \(x = \beta \hbar \omega_0\) = \(0.95\) (red), \(0.975\) (yellow), \(1.025\) (green), and \(1.05\) (blue). We have summed over the first 500 terms of the series appearing in Eq. (\ref{I2final1}).}
\label{fig222}
\end{center}
\end{figure}

Let us now discuss the behavior of \(I_1\) and \(I_2\) appearing in the generalized virial theorem. The expression for \(I_2\) depends upon three characteristic frequency scales, set by \((\hbar\beta)^{-1}\), \(\gamma\), and \(\omega_0\), so we define two dimensionless ratios \(x = \hbar \beta \omega_0\) and \(\rho = \gamma/\omega_0\), such that \(I_2\) reads
\begin{widetext}
\begin{eqnarray}
\beta I_2 = \rho^2x^2 \sum_{n=1}^\infty 4 \pi^2 n^2 \Bigg[ \frac{4}{\big(2x^2 + 8 \pi^2 n^2  - \rho^2 x^2 \big)^2 -\rho^2 (\rho^2 - 4) } -  \frac{1}{(x^2 + 4 \pi^2 n^2)^2 - 4 \pi^2 n^2 \rho^2 x^2} \Bigg]. \label{I2final1}
\end{eqnarray}
\end{widetext}
Clearly, one has \(\beta I_2 \rightarrow 0\) for \(\rho \rightarrow 0\) or \(x \rightarrow 0\). The quantity \(\beta I_2\) has been plotted in figure-(\ref{fig111}), as a function of \(x\) with various values of \(\rho\). Interestingly, independent of the value of \(\rho\) (damping strength) taken, \(\beta I_2\) assumes negative values for \(x < 1\), crosses zero exactly at \(x = 1\), and then increases monotonically with \(x\) for \(x > 1\). Moreover, in figure-(\ref{fig222}), we have plotted \(\beta I_2\) as a function of \(\rho\) for a few selected values of \(x\). The magnitude of \(\beta I_2\) is found to monotonically increase with the increase in the damping strength, while its sign is negative or positive depending upon whether \(x < 1\) or \( x > 1\). In a sense therefore, the parameter \(x\), which happens to be the ratio between the quantum oscillator energy scale \(\hbar \omega_0\) and the thermal energy scale \(k_B T\), appears to be an important control parameter in the sense that one can fine-tune its value (independent of dissipation strength) to control the sign of \(I_2\). \\

Let us comment on the weak-coupling limit for which \(I_{1,2} \rightarrow 0\). This tells us that the averaged kinetic and potential energies of the oscillator are equal, i.e. 
\begin{equation}
\langle K\rangle = \langle V\rangle = \frac{\hbar \omega_0}{4} \coth \bigg(\frac{\hbar \omega_0}{2 k_B T}\bigg),
\end{equation} a result that is well known from elementary statistical mechanics. In other words, the bath-induced contributions to the generalized virial theorem investigated in this paper vanish in this limit, even for \(\hbar \neq 0 \). This appears intriguing but can be understood as follows. It is the non-Markovian nature of the quantum noise that is responsible for non-trivial values of \(I_1\) and \(I_2\). While for Ohmic dissipation, the limit \(\hbar \rightarrow 0\) indeed gives a \(\delta\)-correlated noise from Eq. (\ref{symmetricnoisecorrelation1}), one may also alternatively recover an approximately Markovian noise for \(\hbar \neq 0 \) if one considers the weak-coupling limit. It is because in the weak-coupling limit, up to a first approximation, only those degrees of freedom of the bath whose frequencies are in resonance with the characteristic frequency of the system can be considered relevant. This means we may drop the off-resonant contributions, and approximately write Eq. (\ref{symmetricnoisecorrelation1}) as
\begin{equation}\label{ffff}
\frac{\langle \{ F(t), F(t') \} \rangle}{2} \approx \bigg[ m \gamma \hbar \omega_0 \coth \bigg(\frac{\hbar \omega_0}{2k_B T}\bigg) \bigg] \delta(t-t'),
\end{equation} where we have used the fact that the system's characteristic frequency is given by \(\omega_0\). Quite remarkably, Eq. (\ref{ffff}) tells us that the noise is approximately Markovian, justifying \(I_{1,2} \approx 0\). From Eq. (\ref{ffff}), one may define a `quantum' diffusion constant: \(D =m \gamma \hbar \omega_0 \coth \big(\frac{\hbar \omega_0}{2k_B T}\big)\) which, for \(\hbar \rightarrow 0\) appropriately reduces to \(D \rightarrow 2 m \gamma k_B T\). The same physical situation arises from the quantum master equation in the Born-Markov approximation, wherein all stationary state results are identical to those obtained from the quantum Langevin equation for \(\gamma \rightarrow 0\) \cite{GSA}. Thus, it is precisely the non-Markovian character of the quantum noise, that leads to non-trivial expressions for \(I_1\) and \(I_2\) in any case. \\

\begin{figure}
\begin{center}
\includegraphics[scale=0.61]{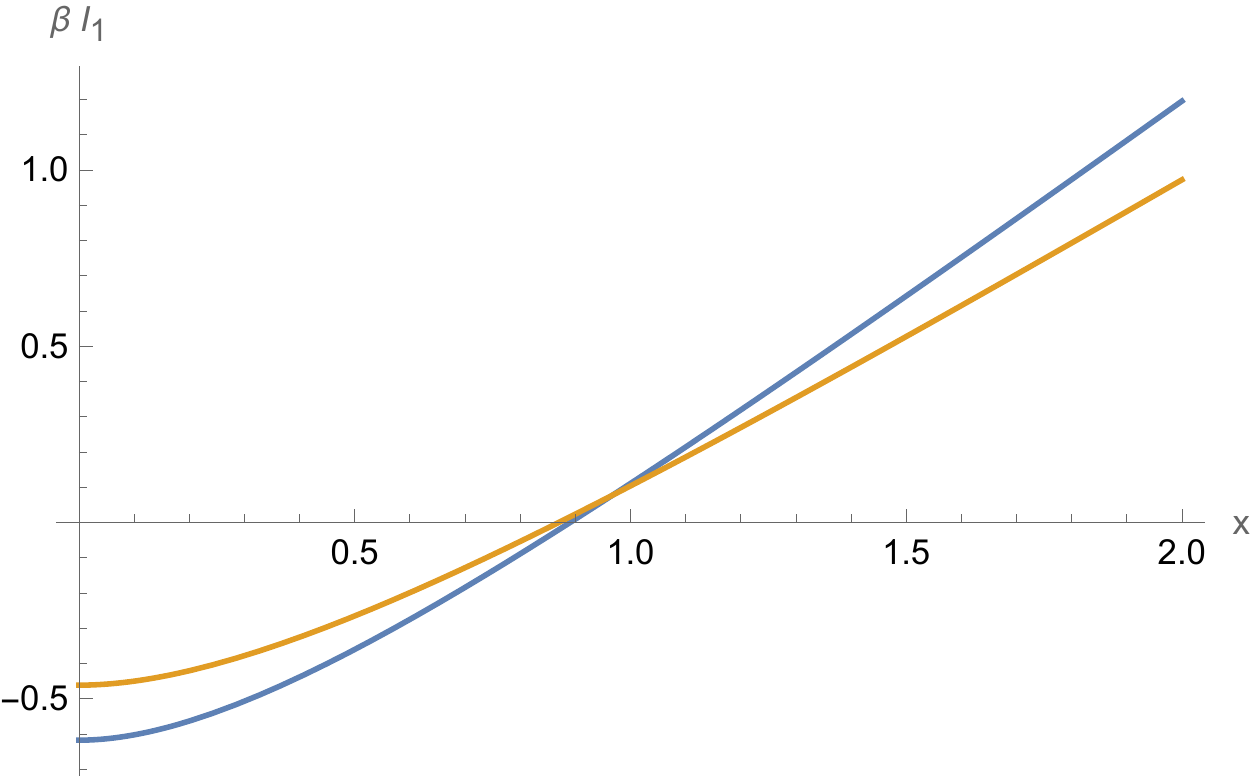}
\caption{Plot of \(\beta I_1\) for the case of Drude dissipation as a function of \(x = \beta \hbar \omega_0\), for \(\rho = \gamma \omega_0^{-1} = 2.1\) (yellow) and \(3.5\) (blue), with \(\sigma = \omega_{\rm cut} \omega_0^{-1} = 10\). We have summed over the first 500 terms of the series appearing in Eq. (\ref{I1drude1}).}
\label{fig2222}
\end{center}
\end{figure}

Coming to \(I_1\), which is physically the average work done by the quantum noise on the particle, we find that it diverges. However, this divergence is not a general feature but is rather tied to the simple case of Ohmic dissipation, which is being considered presently. In order to demonstrate that, let us consider the single-relaxation model, also known as Drude dissipation, for which the bath spectral function assumes a Lorentzian cut-off, i.e. it reads
\begin{equation}
J(\omega) = \frac{m \gamma \omega}{1 + (\omega/\omega_{\rm cut})^2},
\end{equation} where \(\omega_{\rm cut}\) is the cut-off frequency, regularizing the bath spectral function. Notice that \(\tilde{\mu}(\omega)\) and subsequently, \(\alpha(\omega)\) can be computed readily, using the Eqs. (\ref{2}), (\ref{mu}), (\ref{alphadef}), and (\ref{tildemu}). As with the case of Ohmic dissipation, while performing the integral, one has to choose among the cases: \(\frac{\gamma}{2} < \omega_0\) (under-damped regime), \(\frac{\gamma}{2} > \omega_0\) (over-damped regime), or \(\frac{\gamma}{2} = \omega_0\) (critically-damped case). For illustrative purpose, let us choose the over-damped regime. Then \(I_1\) reads
\begin{widetext}
\begin{eqnarray}
I_1 &=& \bigg[\frac{2 \gamma[ \omega_0^2 +  (\omega^2_{\rm cut} - \omega_{\rm cut} \gamma)]}{\beta}\bigg]  \label{I1drude} \\
\times &\Bigg\{ & \frac{1}{(\omega_{\rm cut} + \lambda_1)(\lambda_2 - \lambda_1)(\lambda_1 - \lambda_3)} + \frac{1}{(\omega_{\rm cut} + \lambda_2)(\lambda_2 - \lambda_1)(\lambda_3 - \lambda_2)} + \frac{1}{(\omega_{\rm cut} + \lambda_3)(\lambda_1 - \lambda_3)(\lambda_3 - \lambda_2)} \nonumber \\ 
&+&  \sum_{n=1}^\infty \frac{\nu_n}{(\lambda_1 - \nu_n)(\lambda_2 - \nu_n)(\lambda_3 - \nu_n)(\omega_{\rm cut} + \nu_n)} + \sum_{n=1}^\infty \frac{2 \lambda_1^2}{(\nu_n^2 - \lambda_1^2)(\lambda_2 - \lambda_1)(\lambda_3 - \lambda_1)(\omega_{\rm cut} + \lambda_1)} \nonumber \\
&+& \sum_{n=1}^\infty \frac{2 \lambda_2^2}{(\nu_n^2 - \lambda_2^2)(\lambda_1 - \lambda_2)(\lambda_3 - \lambda_2)(\omega_{\rm cut} + \lambda_2)} + \sum_{n=1}^\infty \frac{2 \lambda_3^2}{(\nu_n^2 - \lambda_3^2)(\lambda_1 - \lambda_3)(\lambda_2 - \lambda_3)(\omega_{\rm cut} + \lambda_3)}  \Bigg\},  \nonumber 
\end{eqnarray}
\end{widetext} where
\begin{equation}
\lambda_{1} = \omega_{\rm cut} - \gamma, \hspace{5mm} \lambda_{2,3} = \frac{\gamma}{2} \pm \sqrt{\frac{\gamma^2}{4} - \omega_0^2} .
\end{equation} 
The series converges in general, demonstrating that the average work done by the quantum noise on the system is finite, and the infinite answer encountered earlier was just a result of choosing Ohmic dissipation. One can observe that for \(\omega_{\rm cut} \rightarrow \infty\), i.e. in the Ohmic limit, Eq. (\ref{I1drude}) reduces to Eq. (\ref{I1final}) and in that case, the series diverges. The series can be expressed in dimensionless form as presented in appendix-(\ref{appB}), by introducing the parameters \(\rho = \gamma/\omega_0\), \(x = \beta \hbar \omega_0\), and \(\sigma = \omega_{\rm cut}/\omega_0\). In figure-(\ref{fig2222}), we have plotted \(I_1\) in units of \(k_BT\) as a function of \(x\), for \(\rho= 2.1\) and \(3.5\), with \(\sigma= 10\). It is clearly seen that \(\beta I_1\) increases with \(x\) and its magnitude increases with \(\rho\). As with \(I_2\), the parameter \(x\) can be tuned to control the sign of \(I_1\).\\

To summarize, we have investigated a generalized virial theorem for the dissipative oscillator in this paper. Two bath-induced corrections were found, where the first one, i.e. \(I_1\) is physically the average work done on the system by the random force on it due to the heat bath, while the second one, i.e. \(I_2\) appears to be related to the average rate of power transfer due to dissipation. It was found that the non-Markovian nature of the bath-induced quantum noise is primarily responsible for the terms \(I_1\) and \(I_2\) appearing in the generalized virial theorem. Therefore, these corrections vanish either in the classical limit, where the noise becomes Markovian, or in the weak-coupling limit. We end by noting that although our general framework is applicable to an arbitrary dissipation mechanism describing a passive heat bath, many of our results were derived keeping the simple and analytically tractable case of Ohmic dissipation in mind. Since the average work done due to the random force on the system diverges for Ohmic dissipation, we computed it for Drude dissipation and demonstrated that it is finite. The generalized virial theorem and its implications should further be investigated for other kinds of baths, including the Drude bath, as well as for different system-bath coupling schemes, most notably momentum-momentum coupling (see \cite{momentum} and references therein). We keep these issues for future work. 

\section*{Acknowledgements}
The authors are grateful to Akash Sinha and Jasleen Kaur for help in preparing the plots. A.G. thanks Sushanta Dattagupta for some clarifying remarks on the Born-Markov approximation of the quantum master equation. The work of A.G. is supported by the Ministry of Education (MoE), Government of India in the form of a Prime Minister's Research Fellowship (ID: 1200454). M.B. is supported by the Department of Science and Technology (DST), Government of India under the Core grant (Project No. CRG/2020/001768) and MATRICS grant (Project no. MTR/2021/000566). We thank Jasleen Kaur for carefully reading the manuscript. 

\begin{widetext}

\appendix

\section{Position-velocity average \(\langle x \dot{x} \rangle\) for classical oscillator systems}\label{appA}
We now show that for the classical Brownian oscillator whose equation of motion is given by Eq. (\ref{classlang}), we have \(\langle x \dot{x} \rangle = 0\), or more explicitly, the two-point correlation function \(\langle x(t) v(t) \rangle = 0\) where \(v(t) = \dot{x}(t)\). We assume that the system is in the steady state, i.e. the Brownian particle has equilibrated with the surrounding fluid. In
 that case, the processes $x(t)$ and $\dot{x}(t)$ are stationary
 random processes~\cite{Zwanzig,balki}, meaning that any two-point correlation
 function involving these is essentially time-homogeneous, i.e.~it
 depends only upon the difference between the two time instants:
 $t-t'$. Notice that the noise is already a stationary process by
 definition. We define the Fourier amplitude of a generic stationary process
 $A(t)$ as
\begin{equation}
\tilde{A}(\omega,\mathcal{T}) = \int_0^\mathcal{T} A(t) e^{i\omega t} d\omega.
\end{equation} From Eq. (\ref{classlang}), the Fourier amplitudes of \(x(t)\) and \(F(t)\) satisfy the following algebraic equation (we put \(\mu = m \gamma\)):
\begin{equation}
m(\omega_0^2 - \omega^2 - i \gamma \omega) \tilde{x}(\omega,\mathcal{T}) = \tilde{F}(\omega,\mathcal{T}),
\end{equation} which means \(\tilde{x}(\omega,\mathcal{T}) = \mathcal{A}_x (\omega) \tilde{F}(\omega,\mathcal{T})\), with
\begin{equation}
\mathcal{A}_x (\omega) = \frac{1}{m(\omega_0^2 - \omega^2 - i \gamma \omega)}.
\end{equation} The power spectrum is defined as
\begin{equation}
 S_{x}(\omega) =  \lim {}_{\mathcal{T} \rightarrow \infty} \frac{1}{\mathcal{T}} \frac{|\tilde{F}(\omega,\mathcal{T})|^2}{m^2(\omega_0^2-\omega^2)^2 + m^2 \gamma^2 \omega^2 }.
\end{equation} 
 It is known that the power spectrum of the noise is just \cite{Zwanzig,balki} (see Eq. (\ref{mnvcx}) below)
\begin{equation}
  S_{F} (\omega) = \lim {}_{\mathcal{T} \rightarrow \infty} \frac{1}{\mathcal{T}} |\tilde{F}(\omega,\mathcal{T})|^2 = \Gamma,
\end{equation}which is why \(F(t)\) is called `white', i.e. power spectrum has no dependence on frequency. Combining these, we have
\begin{equation}\label{abcdefgh}
  S_{x} (\omega) =   \frac{\Gamma}{m^2(\omega_0^2-\omega^2)^2 + m^2 \gamma^2 \omega^2 }.
\end{equation} 
There is a result called the Wiener-Khintchine theorem \cite{WKref1,WKref2}, which relates the power spectrum with the correlation function by an inverse Fourier transform, i.e. 
\begin{equation}\label{1234567890}
 C_{xx} (t-t') := \langle x(t) x(t') \rangle = \frac{1}{2\pi} \int_{-\infty}^{\infty} d\omega e^{-i \omega (t-t')} S_{x} (\omega).
\end{equation} 
We can find the position-velocity autocorrelation function by differentiating the above expression with respect to \(t'\), which gives 
\begin{equation}\label{123456789011}
 C_{xv} (t-t') := \langle x(t) v(t') \rangle = \frac{i}{2\pi} \int_{-\infty}^{\infty} \omega d\omega e^{-i \omega (t-t')} S_{x} (\omega).
\end{equation} 
Therefore, at the same instant, i.e. putting \(t=t'\), we get
\begin{equation}\label{12345678901111}
 \langle x(t) v(t) \rangle = \frac{i}{2\pi} \int_{-\infty}^{\infty} \omega S_{x} (\omega) d\omega ,
\end{equation} which obviously vanishes because the integrand is an odd function and the integration limits are symmetric about \(\omega = 0\). This is unlike the case with the two-point functions \(\langle x(t) x(t)\rangle\) or \(\langle v(t) v(t) \rangle\) where the integrands are even functions, and respectively lead to the mean potential and kinetic energies.\\

Let us just briefly comment on the result \( S_{F} (\omega) = \Gamma\), which was used in the preceding computation. Since Wiener-Khintchine theorem relates the power spectrum with the correlation function, one can write
\begin{equation}\label{mnvcx}
S_{F} (\omega) = \int_{-\infty}^{\infty} d\omega e^{i \omega (t-t')} \langle F(t) F(t') \rangle,
\end{equation} which upon substituting \(\langle F(t) F(t') \rangle = \Gamma \delta(t-t')\) gives \(S_F(\omega) = \Gamma\). \\

The Brownian oscillator which is being discussed should be contrasted with the case of the damped harmonic oscillator. The equation of motion is given by Eq. (\ref{dampedeqn1}), which unlike Eq. (\ref{classlang}), is a deterministic differential equation. Its solutions are just \(x(t) = A e^{\lambda_+ t} + B e^{\lambda_- t}\), where \(\lambda_\pm\) are the roots of the equation \(m \lambda^2 + \mu \lambda +k  = 0.\) It is now easy to verify that the time average of the quantity \(x \dot{x}\) is non-vanishing, i.e. \(\langle x \dot{x} \rangle_t \neq 0\).\\

\section{\(I_1\) for Drude dissipation in dimensionless form}\label{appB}
In this appendix, we quote the expression of \(I_1\) (for the over-damped regime) appearing in Eq. (\ref{I1drude}) in a dimensionless form as has been used to plot figure-(\ref{fig2222}). Using the parameters \(x = \beta \hbar \omega_0\), \(\rho = \gamma/\omega_0\), and \(\sigma= \omega_{\rm cut}/\omega_0\), we may express Eq. (\ref{I1drude}) as

\begin{eqnarray}
\beta I_1 &=& \big[2 \rho[ 1 +  (\sigma^2 - \sigma \rho)]\big]  \label{I1drude1} \\
\times &\Bigg\{ & \frac{1}{(2\sigma - \rho)\bigg(\frac{3\rho}{2} + \sqrt{\frac{\rho^2}{4} - 1} - \sigma \bigg)\bigg(-\frac{3\rho}{2} + \sqrt{\frac{\rho^2}{4} - 1} + \sigma \bigg)} - \frac{1}{\bigg(\sigma + \frac{\rho}{2} + \sqrt{\frac{\rho^2}{4} - 1} \bigg)\bigg(\frac{3\rho}{2} + \sqrt{\frac{\rho^2}{4} - 1} - \sigma \bigg)\bigg(2 \sqrt{\frac{\rho^2}{4} - 1}  \bigg)} \nonumber \\
&-& \frac{1}{\bigg(\frac{\rho}{2} - \sqrt{\frac{\rho^2}{4} - 1} + \sigma \bigg)\bigg(-\frac{3\rho}{2} + \sqrt{\frac{\rho^2}{4} - 1} + \sigma \bigg)\bigg(2 \sqrt{\frac{\rho^2}{4} - 1}  \bigg)} \nonumber \\ 
&+&  \sum_{n=1}^\infty \frac{\big(\frac{2 \pi n}{x}\big)}{\bigg(\sigma - \rho - \frac{2 n \pi}{x}\bigg)\bigg(\frac{\rho}{2} + \sqrt{\frac{\rho^2}{4} - 1} - \frac{2 \pi n}{x}  \bigg)\bigg(\frac{\rho}{2} - \sqrt{\frac{\rho^2}{4} - 1} - \frac{2 \pi n}{x}  \bigg)\bigg(\sigma + \frac{2 \pi n}{x}\bigg)} \nonumber \\
&+& \sum_{n=1}^\infty \frac{2 (\sigma - \rho)^2}{\bigg(\frac{4 \pi^2 n^2}{x^2} - (\sigma - \rho)^2\bigg)\bigg(\frac{3\rho}{2} + \sqrt{\frac{\rho^2}{4} - 1} - \sigma \bigg)\bigg(\frac{3\rho}{2} - \sqrt{\frac{\rho^2}{4} - 1} - \sigma \bigg)(2\sigma - \rho)} \nonumber \\
&-& \sum_{n=1}^\infty \frac{2 \bigg( \frac{\rho}{2} + \sqrt{\frac{\rho^2}{4} - 1} \bigg)^2}{\bigg(\frac{4 \pi^2 n^2}{x^2} - \big( \frac{\rho}{2} + \sqrt{\frac{\rho^2}{4} - 1} \big)^2\bigg)\bigg( -\frac{3\rho}{2} - \sqrt{\frac{\rho^2}{4} - 1} + \sigma \bigg)\bigg(2 \sqrt{\frac{\rho^2}{4} - 1}  \bigg)\bigg( \sigma + \frac{\rho}{2} + \sqrt{\frac{\rho^2}{4} - 1}\bigg)} \nonumber \\
&+& \sum_{n=1}^\infty \frac{2 \bigg( \frac{\rho}{2} - \sqrt{\frac{\rho^2}{4} - 1} \bigg)^2}{\bigg(\frac{4 \pi^2 n^2}{x^2} - \big( \frac{\rho}{2} - \sqrt{\frac{\rho^2}{4} - 1} \big)^2\bigg)\bigg(-\frac{3\rho}{2} + \sqrt{\frac{\rho^2}{4} - 1} + \sigma \bigg)\bigg(2 \sqrt{\frac{\rho^2}{4} - 1}\bigg)\bigg(\sigma + \frac{\rho}{2} - \sqrt{\frac{\rho^2}{4} - 1}\bigg)}  \Bigg\}.  \nonumber 
\end{eqnarray} 
\end{widetext}

\end{document}